\documentclass[aps,prl,longbibliography,twocolumn,showpacs,preprintnumbers,amsmath,amssymb,superscriptaddress]{revtex4-1}
\usepackage{times}
\usepackage{amsfonts}
\usepackage{mathrsfs}
\usepackage{graphicx}
\usepackage{dcolumn}
\usepackage{bm}
\usepackage{color}
\usepackage{diagbox}
\usepackage[utf8]{inputenc}
\usepackage[T1]{fontenc}
\usepackage{xcolor}
\usepackage{pgf}
\usepackage{subfigure}
\usepackage{longtable}
\usepackage{bbm}
\usepackage{multirow}
\usepackage{framed}
\usepackage{relsize}
\usepackage{soul}
\usepackage{amsthm}
\usepackage[colorlinks,bookmarks=false,citecolor=blue,linkcolor=red,urlcolor=blue]{hyperref}
\usepackage{float}

\definecolor{ZYcolor}{rgb}{0.1,0.5,0.4}

\newcommand{\Rom}[1]{ \uppercase\expandafter{\romannumeral#1}}

\newcommand{\kk}{{\bf{{k}}}}

\newcommand{\0}{{\bf{0}}}

\usepackage{dsfont}

\definecolor{ZZYcolor}{rgb}{0.1,0.5,0.4}

\def\be{\begin{equation}}       \def\ee{\end{equation}}
\def\bea{\begin{eqnarray}}      \def\eea{\end{eqnarray}}

\begin{document}

\title{Topological Superconductivity from Unconventional Band Degeneracy with Conventional Pairing}

\author{Zhongyi Zhang}
\affiliation{Department of Physics, Hong Kong University of Science and Technology, Clear Water Bay, Hong Kong, China}
\affiliation{Beijing National Research Center for Condensed Matter Physics, and Institute of Physics, Chinese Academy of Sciences, Beijing 100190, China}
\affiliation{University of Chinese Academy of Sciences, Beijing 100049, China}

\author{Zhenfei Wu}
\affiliation{Department of Physics, University of Florida, Gainesville, Florida, 32601, USA}

\author{Chen Fang}
\affiliation{Beijing National Research Center for Condensed Matter Physics, and Institute of Physics, Chinese Academy of Sciences, Beijing 100190, China}
\affiliation{Kavli Institute for Theoretical Sciences and CAS Center for Excellence in Topological Quantum Computation, University of Chinese Academy of Sciences, Beijing 100190, China}

\author{Fu-chun Zhang}
\affiliation{Kavli Institute for Theoretical Sciences and CAS Center for Excellence in Topological Quantum Computation, University of Chinese Academy of Sciences, Beijing 100190, China}
\affiliation{University of Chinese Academy of Sciences, Beijing 100049, China}
\affiliation{Collaborative Innovation Center for Advanced Microstructure, Nanjing University, Nanjing 210093, China}

\author{Jiangping Hu}
\affiliation{Beijing National Research Center for Condensed Matter Physics, and Institute of Physics, Chinese Academy of Sciences, Beijing 100190, China}
\affiliation{Kavli Institute for Theoretical Sciences and CAS Center for Excellence in Topological Quantum Computation, University of Chinese Academy of Sciences, Beijing 100190, China}

\author{Yuxuan Wang}\email{yuxuan.wang@ufl.edu}
\affiliation{Department of Physics, University of Florida, Gainesville, Florida, 32601, USA}

\author{Shengshan Qin}\email{qinshengshan@bit.edu.cn}
\affiliation{School of Physics, Beijing Institute of Technology, Beijing 100081, China}

\begin{abstract}
We present a new scheme for Majorana modes in systems with nonsymmporhic-symmetry-protected band degeneracy. We reveal that when the gapless fermionic excitations are encoded with conventional superconductivity and magnetism, which can be intrinsic or induced by proximity effect, topological superconductivity and Majorana modes can be obtained.
We illustrate this outcome in a system which respects the space group $P4/nmm$ and features a fourfold-degenerate fermionic mode at ($\pi$, $\pi$) in the Brillouin zone.
We show that in the presence of conventional superconductivity, different types of topological superconductivity, i.e. first-order and second-order topological superconductivity, with coexisting fragile Wannier obstruction in the latter case, can be generated in accordance with the different types of magnetic orders; Majorana modes are shown to exist on the boundary, at the corner and in the vortices.
To further demonstrate the effectiveness of our approach, another example related to the space group $P4/ncc$ based on this scheme is also provided.
Our study offers insights into constructing topological superconductors based on bulk energy bands and conventional superconductivity, and helps to find new material candidates and design new platforms for realizing Majorana modes.
\end{abstract}

\maketitle
\section{Introduction}
Topological superconductors \cite{RevModPhys.82.3045,RevModPhys.83.1057,RevModPhys.88.035005,Alicea_2012,classification1} (TSCs) are renowned for hosting a special kind of quasiparticles, the Majorana modes, whose antiparticles are themselves.
Owing to their potential application in the fault-tolerant quantum computation \cite{lian2018topological,Alicea_2012, Kitaev_2001}, a substantial effort has been made to search for the Majorana modes, and great advances have been achieved both in theory \cite{PhysRevLett.100.096407, PhysRevLett.105.097001, PhysRevLett.104.040502, PhysRevLett.105.077001, PhysRevLett.111.056402, PhysRevLett.115.127003, PhysRevX.9.011033,  PhysRevLett.112.106401, PhysRevLett.123.027003,classification1,PhysRevResearch.2.043300,Wang_toappear} and in experiment \cite{TSC_Ando,  nadj2014observation, MZM_Jia, wang2018evidence, vortex_Feng, vortex_Kong, machida2019zero, kong2020tunable, MZM_flux, Majorana_lattice} over the past few decades.
The $p$-wave superconductors have been suggested as promising candidates for the TSCs, and experimental signatures of $p$-wave superconductivity has been detected \cite{RevModPhys.74.235, UTe2, KCrAs_zheng}.
Various artificial devices have been proposed to support topological superconductivity, such as the heterostructure between a conventional superconductor and a topological insulator \cite{PhysRevLett.100.096407} or the Rashba electron gas \cite{PhysRevLett.104.040502, PhysRevLett.105.077001}, and experimental evidences for the Majorana modes have been observed \cite{MZM_Jia, MZM_flux}.
Despite the progress, an efficient way towards platforms realizing the numerous exotic topological superconducting phases \cite{ real_song, TSC_ZYZhang, conventional_watanabe, benalcazar2017quantized, PhysRevLett.119.246402, schindler2018higher,  PhysRevB.97.205135, Zhang_RKKY,ono2023towards,PhysRevLett.123.167001,rotation_CFang,PhysRevLett.112.106401,PhysRevB.98.245413,PhysRevB.99.195431,PhysRevLett.123.156801,PhysRevResearch.2.012060,wong2023higher,zou2021new}, especially the high-order topological superconducting states, is still elusive.

\begin{figure}[!htbp]
	\centering
	\includegraphics[width=.8\linewidth]{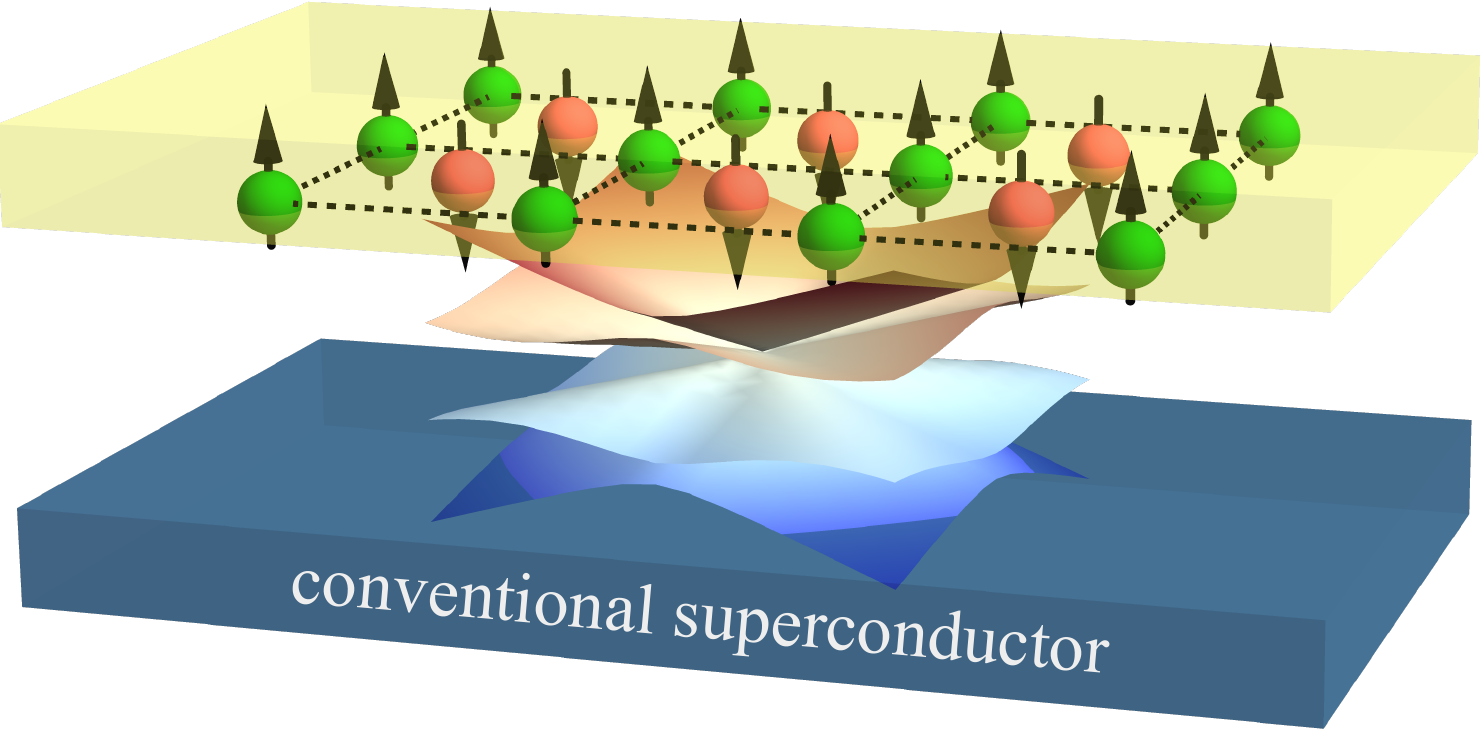}
	\caption{\label{fig1} A sketch for platforms realizing topological superconductivity.
  It is based on systems with gapless fermionic excitations (the intermediate layer) protected by nonsymmorphic crystal symmetries.
    In the system, the magnetism (top) and the conventional superconductivity (bottom) can be induced through either the proximity effect or the intrinsic properties of the intermediate layer.
    The colored balls, black arrows and colored cones represent the different lattice sites, the magnetic moments and energy dispersion, respectively.}
\end{figure}

In recent years, remarkable strides have been made in understanding the topological states of matter.
It is realized that, the topological property of a system can be indicated by the symmetry information of its occupied bands at high-symmetry points, and the system must be topologically nontrivial if its symmetry information at these points differs from that of an atomic insulator \cite{bradlyn2017topological, po2017symmetry, song2018quantitative}.
A parallel formalism has also been developed for the TSCs \cite{doi:10.1126/sciadv.aaz8367, PhysRevB.101.245128, PhysRevResearch.3.023086}.
Motivated by these achievements, we suggest a new scheme to realize TSCs, built on the heterostructure sketched in Fig.~\ref{fig1}, based on symmetry-protected band degeneracies near the fermi energy and conventional superconductivity.
Such fermionic modes, i.e. the band degeneracies, always carry different quantum numbers, such as rotation eigenvalues, mirror eigenvalues, etc.
To make the core of our proposal clearer, let us start with the time-reversal symmetric BdG Hamiltonian with conventional superconductivity, i.e. the uniform $s$-wave pairing.
In such a system, the chiral symmetry which is the combined operation of the time-reversal symmetry and the particle-hole symmetry, maps a negative-energy state to a positive-energy state; Moreover, the unitary chiral symmetry commutes with the crystalline symmetries \cite{RevModPhys.88.035005, TSC_ZYZhang,rotation_CFang}, leading to that the two states related by the chiral symmetry carry the same quantum numbers. This property implies that in the system, the information of the symmetry eigenvalues corresponding to all the negative-energy states at the high-symmetry point in the Brillouin zone are always the same as the condition where the normal-state electronic states are fully occupied or fully unoccupied, which must be topologically trivial \cite{doi:10.1126/sciadv.aaz8367}. Notice that the above conclusion is always true, regardless of the location of the Fermi energy.
Therefore, in the sense of the symmetry indicator, any time-reversal symmetric superconductor with uniform $s$-wave pairing is topologically trivial~\cite{doi:10.1126/sciadv.aaz8367, PhysRevB.101.245128, PhysRevResearch.3.023086}.
However, if the time-reversal symmetry is broken such as by the magnetic orders, the above symmetry constraint fails. Moreover, as long as the eigenvalues of the crystalline symmetries carried by the negative-energy states are different from that in the time-reversal symmetric case, some nontrivial topology is indicated in the superconductor; And such a condition is most likely to occur, when there is band degeneracy near the Fermi energy.
More specifically, when the band degeneracy is encoded with magnetism, it will split; If the chemical potential resides within the split band gap, in the superconducting state the symmetry eigenvalues carried by the positive-energy states will no longer match those of the negative-energy states, indicating the presence of nontrivial topology (more details in Supplementary Note 1).
We illustrate this scheme in a system respecting the space group $P4/nmm$ and show various topological superconducting states can be achieved in accordance with the different magnetic orders.
To further show the effectiveness of our approach, we provide another example related to the space group $P4/ncc$ in the Supplementary Note 10.
Compared with earlier proposals~\cite{PhysRevLett.104.040502,Alicea_2012,mourik2012signatures,PhysRevLett.105.077001}, the key advantange here is that by leveraging the nonsymmorphic crystalline symmetries, the resulting phases of topological superconductivity is much richer.
In recent years, the distinct irreducible representations (IRs) of the little group of the crystalline symmetries can assist in identifying different types of free fermionic excitations, such as the unconventional quasiparticles beyond Dirac and Weyl fermions \cite{bradlyn2016beyond}.
Based on those abundant fermionic excitations, our method can be applied to a wide range of systems, and opens up a new direction of searching for novel topological superconducting phases in these materials.

In the following, we focus on the space group $P4/nmm$ which has a four-dimensional irreducible projective representation at the Brillouin zone corner. We show that the antiferromagnetic (AFM) order and ferromagnetic (FM) order can both split the fourfold degeneracy into two twofold ones.
In the presence of conventional superconductivity, the AFM order drives the system into a second-order TSC state coexisting with fragile Wannier obstruction, while the FM order results in a first-order TSC, as long as the chemical potential lies in the magnetic gap. These results may be relevant to iron-based superconductors and heterostructures thereof, which hosts intrinsic AFM order and high-$T_c$ superconductivity.

\section{Results}
\subsection{Fourfold degenerate fermion with SG 129.}
\begin{figure}[t]
	\centering
	\includegraphics[width=1\linewidth]{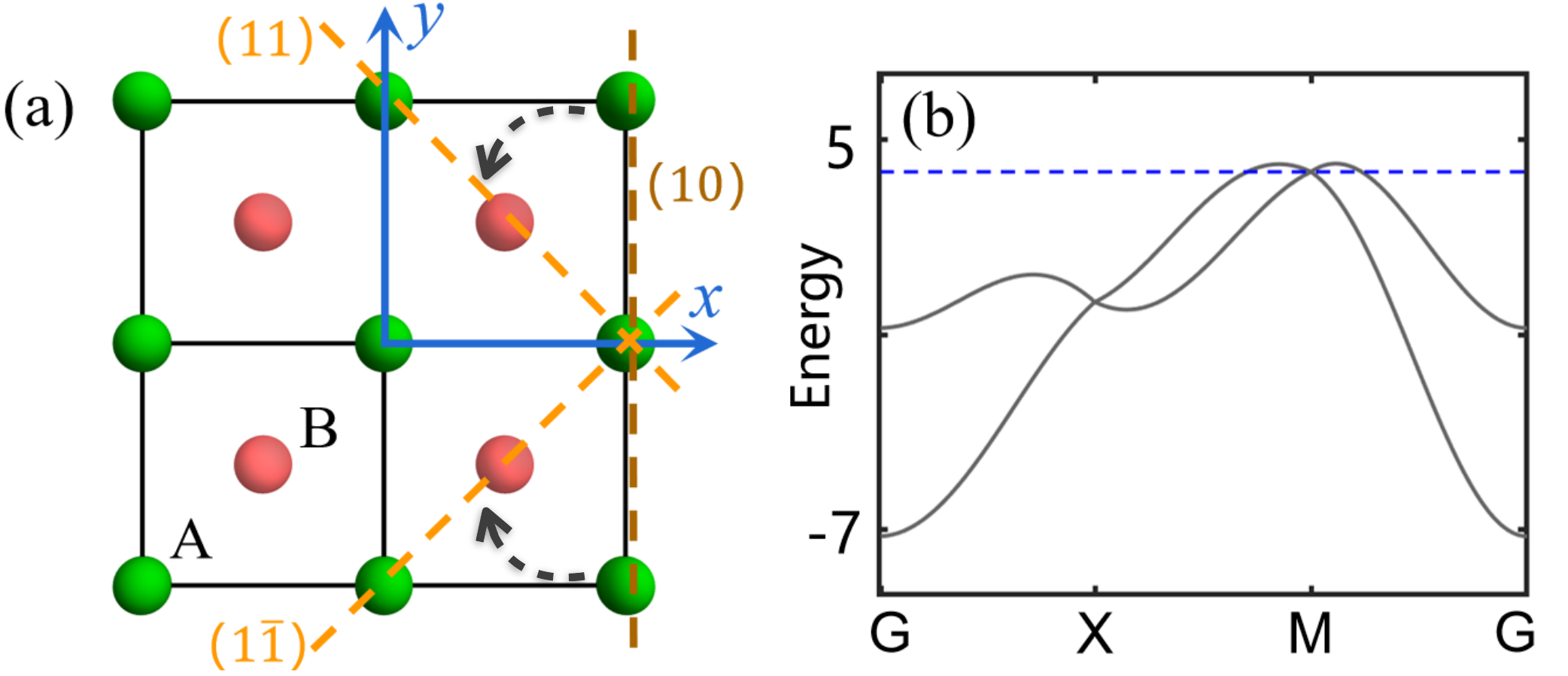}
	\caption{\label{fig2} Lattice structure and band sturcture of paramagnetic state.
  (a) shows a quasi-2D lattice respecting group $P4/nmm$. The green and red balls label the two sublattices. The orange and brown dashed lines indicate the different edges considered in the text. The black dashed arrows represent a bending process from the $(10)$ edge to the $[11]$ and $[1 \overline{1} ]$ edges.
    (b) presents the bands obtained from Eq.~\eqref{normal_Hamiltonian}, with G, X, M representing $(0, 0)$, $(\pi, 0)$, $(\pi, \pi)$ in the Brillouin zone respectively, with the other parameters are set to be $\{ t, t^\prime, \lambda \} = \{ -1.0, 0.8, 0.5\}$. The blue dashed line in (b) represents the chemical potential considered in the text. }
\end{figure}

We begin with an introduction of the space group $\mathcal{G} = P4/nmm$ ($\#.129$), i.e. the symmetry group governing the iron-based superconductors.
We focus on the quasi-two dimensional (2D) case, and consider the lattice in Fig.~\ref{fig2}(a) which is similar to the monolayer FeSe.
The space group $P4/nmm$ is nonsymmorphic, and it has a special group structure as follows \cite{PhysRevX.3.031004}
\begin{align}\label{quotientG_main}
\begin{split}
\mathcal{G}/T &= D_{2d} \otimes Z_2,
\end{split}
\end{align}
where $T$ is the translation group, $D_{2d}$ is the point group at the lattice sites, and $Z_2$ is a two-element group including the inversion symmetry which switches the two sublattices in the lattice in Fig.~\ref{fig2}(a). As $D_{2d}$ and $Z_2$ are defined on different points, Eq.~\eqref{quotientG_main} holds in a sense that symmetry operations are equivalent if they differ by a lattice translation, hence the quotient group on the left hand side. According to Eq.~\eqref{quotientG_main}, $\mathcal{G}/T$ can be generated by the generators of $D_{2d}$ and $Z_2$, including the inversion symmetry $\{ I | {\bf \tau}_0 \}$, the mirror symmetry $\{ M_y | {\bf 0} \}$ and the rotoinversion symmetry $\{ S_{4z} | {\bf 0} \}$. Here, we express the symmetry operations in the form of the Seitz operators. In the generators, the point group parts act on the Cartesian coordinates as $I: (x, y, z) \mapsto (-x, -y, -z)$, $M_y: (x, y, z) \mapsto (x, -y, z)$, and $S_{4z}: (x, y, z) \mapsto (y, -x, -z)$, and ${\bf \tau}_0 = {\bf a}_1/2 + {\bf a}_2/2$ with ${\bf a}_1$ (${\bf a}_2$) the primitive lattice translation along the $x$ ($y$) direction in Fig.~\ref{fig2}(a).


For electronic systems in the presence of spin-orbit coupling, group $P4/nmm$ has only one single 4D IR at $(\pi, \pi)$, i.e. the M point in the Brillouin zone, where all the symmetry operations in $\mathcal{G}/T$ are respected.
It describes the fourfold degeneracy composed of two Kramers' doublets $J_z = \pm 1/2$ and $J_z = \pm 3/2$ with opposite parities, where $J_z$ is the angular momentum defined according to $\{ S_{4z} | {\bf 0} \}$.
The degeneracy can be understood from the group structure in Eq.~\eqref{quotientG_main}.
The point group $D_{2d}^D$ (double group version of the point group $D_{2d}$) supports two different 2D IRs corresponding to Kramers' doublet $J_z = \pm 1/2$ and $J_z = \pm 3/2$ separately.
At the M point, $\{ S_{4z} | {\bf 0} \}$ in $D_{2d}^D$ and $\{ I | {\bf \tau}_0 \}$ in $Z_2$ satisfy the following anticommutation relation
\begin{eqnarray}\label{main_129commute}
&& \quad \ \  \{ S_{4z} | {\bf 0} \} \{ I | {\bf \tau_0} \} | \varphi({\bf k}) \rangle
= \{ I | {\bf \tau}_0 \} \{ S_{4z} | {\bf a_2} \} | \varphi({\bf k})    \\
&& = e^{i {\bf k \cdot a_2}} \{ I | {\bf \tau_0} \} \{S_{4z} | {\bf 0} \} | \varphi({\bf k}) \rangle
= -\{ I | {\bf \tau_0} \} \{S_{4z} | {\bf 0} \} | \varphi({\bf k}) \rangle,   \nonumber
\end{eqnarray}
which enforces the degeneracy between the two 2D IRs labeled by $J_z = \pm 1/2$ and $J_z = \pm 3/2$ at M (more detailed analysis in Supplementary Note 2).
In the paramagnetic state, besides crystalline symmetries the time-reversal symmetry $\mathcal{T}$ also exists. Correspondingly, the system actually respects the type-II magnetic space group $P4/nmm 1^\prime$ ($\#.129.412$), which reads
\begin{equation}\label{group_sturcture_PM}
\begin{aligned}
\mathcal{G}_{\text{PM}}/T &= D_{2d}^D \otimes Z_2 \otimes \{ \{ E | {\bf 0} \}, \mathcal{T} \}.
\end{aligned}
\end{equation}
Notice that the time-reversal symmetry does not affect the 4D fermionic IR at M.

Assuming trivial band structure at other high-symmetry points, we describe the fourfold degenerate fermion at M by the following tight-binding model \cite{qin_triplet, PhysRevLett.119.267001}
\begin{eqnarray}\label{normal_Hamiltonian}
\mathcal{H}_0({\bf k}) &=& 2t (\cos k_x + \cos k_y) s_0\sigma_0 + 4t^\prime \cos\frac{k_x}{2} \cos\frac{k_y}{2} s_0\sigma_1  \nonumber \\
& & - 2\lambda \sin k_x s_2\sigma_3 - 2\lambda \sin k_y s_1\sigma_3,
\end{eqnarray}
where a single $s$ orbital is assumed at each site in the lattice in Fig.~\ref{fig2}(a). In Eq.~\eqref{normal_Hamiltonian}, the Pauli matrices $s_i$ and $\sigma_i$ ($i = 1, 2, 3$) stand for the spin and sublattice degrees respectively. $t$ ($t^\prime$) is the nearest-neighbour intrasublattice (intersublattice) hopping. $\lambda$ is the inversion-symmetric Rashba spin-orbit coupling which arises due to the mismatch between the lattice sites and the inversion center \cite{ PhysRevX.12.011030, local_inversion}. The band structure based on $\mathcal{H}_0({\bf k})$ is plotted in Fig.~\ref{fig2}(b). We set the Fermi energy near the fourfold band degeneracy as indicated in Fig.~\ref{fig2}(b), and consider conventional superconductivity in the system. The corresponding BdG Hamiltonian takes the form
\begin{equation}\label{total_BdG}
\begin{aligned}
\mathcal{H}_{\text{BdG}}({\bf k}) = [ \mathcal{H}_0({\bf k}) - \mu ] \kappa_3 + \Delta_{\text{sc}} s_0 \sigma_0 \kappa_1,
\end{aligned}
\end{equation}
in the basis $\psi^\dagger({\bf k}) = ( c^\dagger({\bf k}), is_2\sigma_0 c(-{\bf k}) )$.
In Eq.~\eqref{total_BdG}, the Pauli matrix $\kappa_i$ describes the Nambu spinor, $\mu$ is the chemical potential, and $\Delta_{\text{sc}}$ is the superconducting order parameter. In the superconducting state, the matrix form for the symmetry generators are $\mathcal{I} = s_0 \sigma_1 \kappa_0$, $\mathcal{M}_y = i s_2 \sigma_3 \kappa_0$ and $\mathcal{S}_{4z} = e^{ is_3\pi/4 } \sigma_3 \kappa_0$ \cite{qin_triplet}, where $\mathcal{I}$, $\mathcal{M}_y$ and $\mathcal{S}_{4z}$ correspond to $\{ I | {\bf \tau_0} \}$, $\{ M_y | {\bf 0} \}$, $\{ S_{4z} | {\bf 0} \}$ respectively.
The time-reversal symmetry takes the form $\mathcal{T} = i s_2 \sigma_0 \kappa_0 K$ and the particle-hole symmetry $\mathcal{P} = s_2 \sigma_0 \kappa_2 K$, with $K$ the complex conjugation operation.  It is easy to check that the system described by $\mathcal{H}_{\text{BdG}}$ in Eq.~\eqref{total_BdG} is topologically trivial.

\begin{figure*}[t]
	\centering
	\includegraphics[width=1\linewidth]{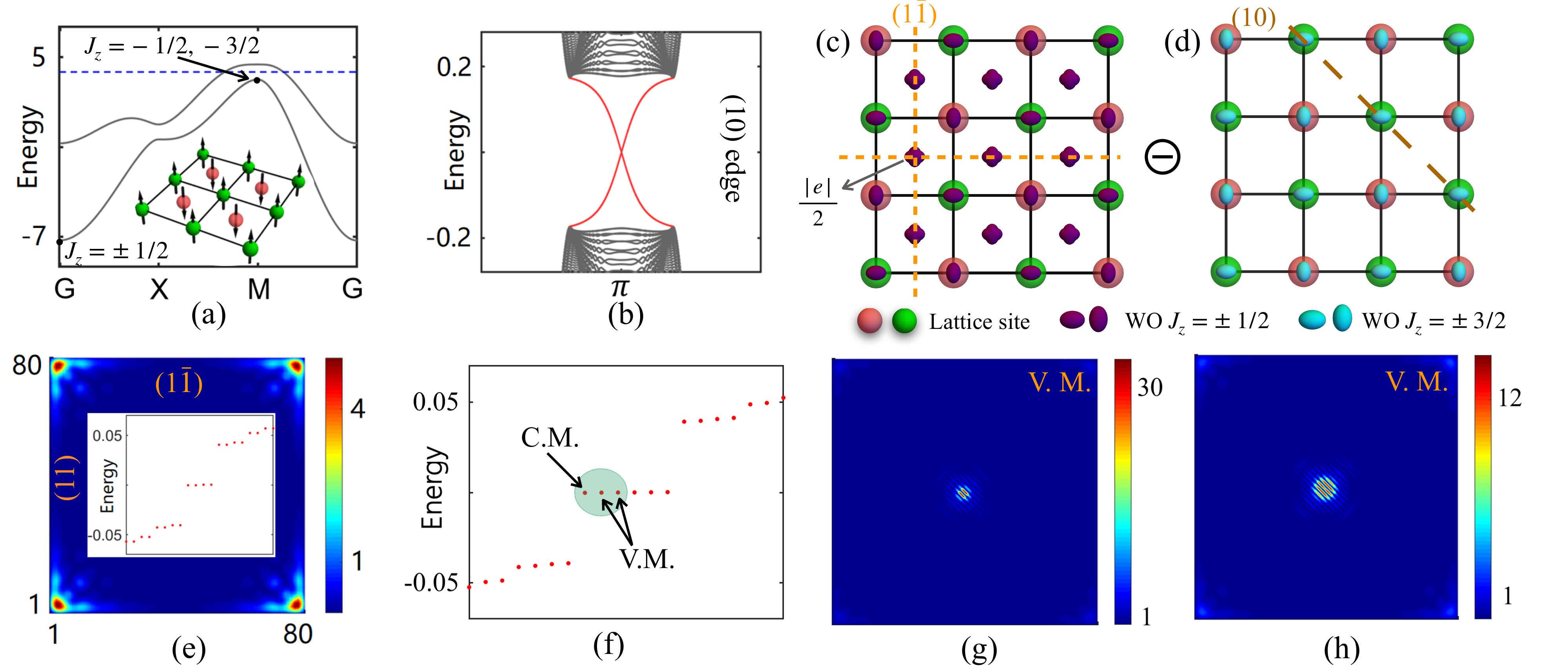}
	\caption{\label{fig3} Distinct manifestations of topology in the AFM case.
  (a) shows the normal bands for the system in Eq.~\eqref{main_AFM} at $\Delta_{\text{AFM}} = 0.5$, with the AFM order illustrated in the inset.
    The blue dashed line represent the chemical potential at $\mu = 4.0$.
     (b) shows the superconducting edge modes corresponding to the bands in (a) on the $(10)$ edge.
     The edge modes on the right and left edges are degenerate.
     (c) shows an atomic insulator constructed by placing two Wannier orbits (WOs) with $J_z=\pm 1/2$ at $2c$ Wyckoff positions (the center of the square formed by the red and green balls), one WO with $J_z=+1/2$ at one of $2a$ Wyckoff positions (red balls) and one WO with $J_z=-1/2$ at the other $2a$ Wyckoff position (green balls).
     (d) shows an atomic insulator constructed by placing one WO with $J_z=+3/2$ at one of $2a$ Wyckoff positions, and one WO with $J_z=-3/2$ at the other $2a$ Wyakoff position.
    (e) shows the low-energy superconducting spectrum (inset) and the real-space wavefunction profiles of the zero-energy modes, corresponding to the bands in (a).
    Open boundary conditions are set in both the $[11]$ and $[1\overline{1}]$ directions.
    (f) shows the low-energy superconducting spectrum in the presence of a single vortex in (e). In the shadow region, among the three zero-energy modes there are two vortex bound Majorana modes (V.M.) and one corner bound Majorana mode (C.M.).
    (g)$\sim$(h) show the real-space wavefunction profiles of the two V.M. in the shadow region in (f), and the C.M. in (f) have similar wavefunction with that in (e). The color bars in  (e)(g)(h) are in the unit of $10^{-3}$.
    In the calculations, the superconducting order is set to be $\Delta_{\text{sc}} = 0.2$.}
\end{figure*}


\subsection{AFM order induced second-order TSCs.}
We study possible topological superconductivity in the structure sketched in Fig.~\ref{fig1}, based on the above fourfold degenerate fermion. First, we consider the checkboard AFM order preserving the translational symmetries in the system as illustrated in Fig.~\ref{fig3}(a), and we assume the magnetic polarization along the $z$ direction. Correspondingly, the system is described by the following Hamiltonian
\begin{equation}\label{main_AFM}
\begin{aligned}
\mathcal{H}_{\text{BdG,AFM}} = \mathcal{H}_{\text{BdG}} + \Delta_{\text{AFM}} s_3 \sigma_3 \kappa_0,
\end{aligned}
\end{equation}
with $\Delta_{\text{AFM}}$ the strength of the AFM order.
It is easy to check that, the system respects the type-III magnetic space group $P4^\prime/n^\prime m^\prime m$ ($\#.129.416$)
\begin{equation}\label{group_sturcture_AFM}
\mathcal{G}_{\text{AFM}}/T = S_4^D \otimes \{ \{ E | {\bf 0} \}, \{ M_{xy} | {\bf \tau}_0 \} \} \otimes \{ \{ E | {\bf 0} \}, \{ I | {\bf \tau}_0 \} \mathcal{T} \}.
\end{equation}
We consider the effect of the AFM order on the fourfold degeneracy at M.
Obviously, all the symmetry operations in $\mathcal{G}_{\text{AFM}}/T$ preserve at the M point.
A direct analysis shows that, the fourfold degeneracy is broken into two twofold degenerate ones.
It is the $J_z = 1/2$ ($J_z = -1/2$) state that is degenerate with the $J_z = 3/2$ ($J_z = -3/2$) state. A detailed group analysis is presented in the Supplementary Note 3. Such twofold band degeneracies arise from the relation $\{ S_{4z} | {\bf 0} \} \{ M_{xy} | {\bf \tau}_0 \} = \{ M_{xy} | {\bf \tau}_0 \} \{ \overline{S}_{4z}^3 | {\bf a_1} \}$, which at M leads to
\begin{equation}\label{commutation_AFM}
\{ S_{4z} | {\bf 0} \} \{ M_{xy} | {\bf \tau}_0 \} | \varphi({\bf k}) \rangle
= -\{ M_{xy} | {\bf \tau}_0 \} \{ \overline{S}_{4z}^3 | {\bf 0} \} | \varphi({\bf k}) \rangle.
\end{equation}
Recalling that $\overline{S}_{4z} = S_{4z}^5$, one immediately comes to the above conclusion. We simulate the bands in the presence of the AFM order numerically, and show the results at $\Delta_{\text{AFM}} = 0.5$ in Fig.~\ref{fig3}(a). Here, it is worth mentioning that the bands in Fig.~\ref{fig3}(a) are always twofold degenerate due to the symmetry $\{ I | {\bf \tau_0} \} \mathcal{T}$ which exists at every ${\bf k}$ point in the Brillouin zone and satisfies $( \{ I | {\bf \tau_0} \} \mathcal{T} )^2 = -1$.

As the magnetism breaks the time-reversal symmetry but preserves the particle-hole symmetry, the system belongs to class D which in the 2D case is characterized by a $\mathbbm{Z}$ topological index, i.e. the Chern number, according to the Altland-Zirnbauer classification \cite{classification1}.
The Chern number can be calculated efficiently based on the symmetry eigenvalues carried by the occupied bands at the high-symmetry points. In systems respecting the fourfold rotational symmetry $C_{4}$, in the weak-pairing condition the Chern number $C_h$ satisfies \cite{rotation_CFang}
\begin{equation}\label{Chern_formula}
\begin{aligned}
e^{i2\pi C_h/4} = \frac{ \xi^2(\Gamma) }{ \xi^2({\text M}) } e^{ -\frac{i2m\pi}{4} [ N_{{\text{occ}}}(\Gamma) + N_{{\text{occ}}}({\text M}) - 2N_{{\text{occ}}}({\text X}) ] },
\end{aligned}
\end{equation}
where $m$ is the angular momentum carried by the Cooper pair, $\xi(\Gamma)$ and $\xi({\text M})$ are the products of the $C_{4}$ eigenvalues of the occupied bands at $\Gamma$ and M respectively, and $N_{{\text{occ}}}(\Gamma)$, $N_{{\text{occ}}}({\text M})$ and $N_{{\text{occ}}}({\text X})$ are the number of the occupied bands at $\Gamma$, M and X respectively. Since $C_{4}$ is equivalent to $S_{4}$ in 2D systems, the formula in Eq.~\eqref{Chern_formula} applies to our consideration \cite{footnote2}. The conventional superconductivity carries zero angular momentum, i.e. $m = 0$.
Therefore, the Chern number is completely determined by the $S_{4z}$ eigenvalues of the occupied bands at $\Gamma$ and M, and for the condition in Fig.~\ref{fig3}(a) we find that $C_h = 0$, which is also confirmed by the gapped modes on the $(11)$ and $[1\overline{1}]$ edges (see Supplementary Note 5).
Nonetheless, the system is topologically nontrivial as evidenced by the helical edge mode on the $(10)$ edge in Fig.~\ref{fig3}(b).
In fact, the system is a TSC protected by the antiunitary symmetry $\mathcal{M}_y \mathcal{T}$.
We focus on high-symmetry line $k_y = \pi$, where $\mathcal{M}_y \mathcal{T}$ and the particle-hole symmetry $\mathcal{P}$ preserve. Moreover, $\mathcal{M}_y \mathcal{T}$ serves as a pseudo time-reversal symmetry on line $k_y = \pi$ satisfying $( \mathcal{M}_y \mathcal{T} )^2 = 1$. Therefore, the $k_y = \pi$ line can be viewed as a 1D subsystem of the whole system, which belongs to symmetry class BDI \cite{classification1}. The topological property of such a system is featured by the winding number,
\begin{align}\label{main_winding}
\begin{split}
w &= \int_{-\pi}^{\pi} \frac{dk_x}{2\pi}\ \mathrm{Tr}[ \widetilde{\mathcal{C}} \mathcal{H}_{\text{BdG,AFM}}^{-1}({\bf k}) \partial_{k_x} \mathcal{H}_{\text{BdG,AFM}}({\bf k}) ],
\end{split}
\end{align}
with $\widetilde{\mathcal{C}} = \mathcal{M}_y \mathcal{T} \mathcal{P}$ being the pseudo chiral symmetry on $k_y = \pi$.
We calculate the winding number straightforwardly and it turns out $w = 2$ (details in Supplementary Note 4), which is consistent with the two zero energy modes at $k_y = \pi$ on the $(10)$ edge presented in Fig.~\ref{fig3}(e).

More interestingly, the above even winding number state is actually a second-order TSC state \cite{benalcazar2017quantized, PhysRevLett.119.246402, schindler2018higher, PhysRevB.97.205135} protected by $\mathcal{M}_y \mathcal{T}$.
We demonstrate it numerically.
As presented in Fig.~\ref{fig3}(e), a single Majorana mode exists at the corner between the neighbouring $(11)$ and $(1 \overline{1} )$ edges.
To understand the phenomenon, we start with the helical mode in Fig.~\ref{fig3}(b).
On the $(10)$ edge, the symmetry $\{ M_y | {\bf 0} \} \mathcal{T}$ and the particle-hole symmetry preserve. Considering the two symmetries, we can get the effective theory on the $(10)$ edge as $\mathcal{H}_{(10)} = v k_y \eta_1$, with $v$ the Fermi velocity and $\eta_i$ the Pauli matrices in the space spanned by the helical edge mode. Then, we bend edge $(10)$ into a right angle, with the two sides along the $[11]$ and $[1 \overline{1} ]$ directions as illustrated in Fig.~\ref{fig2}(a).
The helical mode on each edge gains a mass, since $\{ M_y | {\bf 0} \} \mathcal{T}$ breaks on the $(11)$/$(1 \overline{1} )$ edge. The gapped edge modes are depicted by the following effective theory
\begin{equation}\label{edge}
\begin{aligned}
\mathcal{H}_{(11)} = v k \eta_1 + m_{ (11) } \eta_3,  \ \
\mathcal{H}_{(1 \overline{1} )} = v k \eta_1 + m_{ (1 \overline{1} ) } \eta_3,
\end{aligned}
\end{equation}
where $m_{ (11) / (1 \overline{1} ) }$ is the mass term on the $(11)$/$(1 \overline{1} )$ edge. Moreover, $\{ M_y | {\bf 0} \} \mathcal{T}$ requires $m_{ (11) } = -m_{ (1 \overline{1} ) }$.
Therefore, Eq.~\eqref{edge} describes a massive Dirac theory, with the mass changing sign at the corner between the $(11)$ and $(1 \overline{1} )$ edges. The mass domain results in a single Majorana mode at the corner \cite{ PhysRevLett.110.046404, PhysRevLett.121.096803, PhysRevLett.121.186801}.
Due to the pseudo chiral symmetry $\widetilde{\mathcal{C}}$, the corner Majorana modes carries chirality and
the modes with same chirality cannot hybridize with each other. Thus, the classification for the second-order TSC here is $\mathbbm{Z}$.
Moreover, it is worth pointing out the above second-order TSC state exists in the condition $(4t + \mu)^2 + \Delta_{{\text {sc}}}^2 < \Delta_{{\text{AFM}}}^2$, i.e. the chemical potential in the AFM gap in the weak-pairing condition, and it belongs to a $\mathbbm{Z}$ classification corresponding to the winding number along $k_y = \pi$ protected by $\{ M_y | {\bf 0} \} \mathcal{T}$. We present more detailed analyses on the above effective edge theory and the topological phase transitions in the Supplementary Note 5.

Interestingly, the negetive energy states of the BdG Hamiltonian in Eq.~\eqref{main_AFM} displays both fragile Wannier obstruction and second-order topology. To this end, we treat the BdG band structure as an insulator, i.e., ignoring the particle-hole symmetry. Noting that particle-hole partners in the BdG bands carry opposite angular momenta, the angular momenta of the four ``occupied" (negative energy) BdG bands are $J_z =\pm 1/2,\pm 1/2$ at G, $J_z = -1/2, -1/2, -3/2, -3/2$ at M, and $J_z=\pm 1/2, \pm 1/2$ at X. By exhaustion, one can show that no Wannier representation exist. However, if one includes two additional trivial bands (e.g., from core electrons) that are equivalent with two Wannier orbitals with $J_z=\pm 3/2$ each at one of the 2a Wyckoff positions shown in Fig~\ref{fig3}(d), the combined six bands, nevertheless, become Wannier representable. The six Wannier orbitals are centered at Wyckoff position $2c$ with angular momenta $J_z =\pm 1/2,\pm 1/2$ and Wyckoff position $2a$ with $J_z=-1/2, 1/2$, as shown in Fig.~\ref{fig3}(c). Therefore, the occupied bands, despite not Wannier representable, can be viewed as the difference between two Wannier representable systems, with six and two occupied bands respectively,  as shown in Figs.~\ref{fig3}(c) and (d). By definition, the four occupied bands display the fragile Wannier obstruction \cite{PhysRevLett126402}. Formally, using the modern language of magnetic elementary band representation~\cite{nc.MBER}, we express the fragile Wannier obstruction protected by the magnetic space group symmetries in the Supplementary Note 6.

The elucidation of the fragile Wannier obstruction enables an alternative understanding of the second-order topology invoking only $S_{4z}$.
Ignoring the particle-hole symmetry, the stable second-order topology degenerates into the fragile Wannier obstruction.
More specifically, from Fig.~\ref{fig3}(c), the six-orbital Wannier representation displays a filling anomaly. Indeed, viewed as an insulator, if we neglect the difference between two 2a sites in Fig.~\ref{fig3}(c), and combine both the ionic charge and electronic charge at 2a, the configuration is exactly the same as the Benalcazar-Bernevig-Hughes model~\cite{benalcazar2017quantized} for higher-order topology protected by fourfold rotation symmetry (equivalent with our $S_{4z}$), only rotated by 45 degrees.
It can be verified from Ref. \cite{Takahashi20212021} that our model hosts a corner charge $e/2$ because of the mismatch of charge neutrality and rotation symmetry, which ensures a degeneracy of four corner states.
In our system, corner states are pinned at zero energy by the particle-hole symmetry and they are Majorana zero modes. Since the filling anomaly requires only $S_{4z}$, the corner zero modes are stable even when the corner is asymmetric under $\{ M_y | {\bf 0} \}$.
In fact, to reveal the corner charge in a $S_{4z}$ symmetric sample, one only needs to avoid the edge terminations (10) and (01) where gapless edge modes are present due to additional mirror symmetries $\{ M_y | {\bf 0} \}$.
Considering the various topology in the system, for clarity we summarize the relation between the symmetry and the topology in Table.~\ref{Summary table}.

\subsection{Vortex bound Majorana modes.}
In the the second-order TSC state in the above, each vortex can bind two Majorana modes which are stable due to the $S_{4z}$ symmetry. The phenomenon is closely related to the fact that for group $P4/nmm$, the effective theory near M in the normal state can be viewed as a direct sum of two Rashba electron gas systems with angular momenta $J_z = \pm 1/2$ and $J_z = \pm 3/2$ separately.
To make it clearer, we consider the low-energy theory near M in the second-order TSC state for instance
\begin{equation}\label{main_vortex}
\begin{split}
\mathcal{H}_{\mathrm{eff}}({\bf q})& =  [ -t( q_x^2 + q_y^2 ) + 2\lambda ( q_x s_2 \sigma_3 + q_y s_1 \sigma_3 ) ] \kappa_3   \\
&\quad + t^\prime q_x q_y \sigma_1 \kappa_3   + \Delta_{\text{AFM}} s_3 \sigma_3 + \Delta_{\text{sc}} \kappa_1,
\end{split}
\end{equation}
where ${\bf q}$ is defined with respect to the M point, and the identity matrices are omitted for simplicity.
Ignoring the high-order $t^\prime q_x q_y$ term, it is obvious to notice that $\mathcal{H}_{\mathrm{eff}}$ can be decoupled in the $\sigma$ space, i.e. the sublattice space.
In the $\sigma = \pm 1$ subspace, it describes a superconducting Rashba electron gas in the presence of a Zeeman field $\pm\Delta_{\text{AFM}}$; and in each subspace the vortex can bind a single Majorana mode \cite{PhysRevLett.104.040502} carrying $\mathcal{S}_{4z}$ eigenvalue $1$. Notice that in the presence of a vortex, the $S_{4z}$ symmetry takes eigenvalues $\pm 1$ and $\pm i$.
However, the $\sigma = +1$ subspace is spanned by the Kramers' doublet $J_z = \pm 1/2$, while $\sigma = -1$ subspace spanned by $J_z = \mp 3/2$, which can be inferred from the basis of $\mathcal{H}_{\mathrm{eff}}$.
When we consider the $\mathcal{S}_{4z}$ eigenvalue of the Majorana mode, in the $\sigma = -1$ subspace the basis contributes an additional phase factor $e^{i \mp \pi} = -1$.
Therefore, the vortex bound Majorana mode in the $\sigma = \pm 1$ subspace has $\mathcal{S}_{4z}$ eigenvalue $\pm1$.
The two Majorana modes are immune to perturbations preserving the $\mathcal{S}_{4z}$ symmetry, such as the $t^\prime q_x q_y$ term in Eq.~\eqref{main_vortex}. Namely, the second-order TSC state in the above supports two Majorana modes in each vortex protected by the $S_{4z}$ symmetry, i.e. one with $S_{4z}$ eigenvalue $+1$ and the other $-1$.
We carry out numerical simulations for the vortex bound states and present the results in Figs.~\ref{fig3}(f)-(h).
It is interesting to notice that in the second-order TSC state, the corner MZMs in Fig.~\ref{fig3}(e) coexist with the two vortex bound MZMs. This arises from the fact that, the vortex core is far away from the corners, making the corner MZMs can hardly feel the effect of the vortex.

\begin{figure}[b]
	\centering
	\includegraphics[width=1\linewidth]{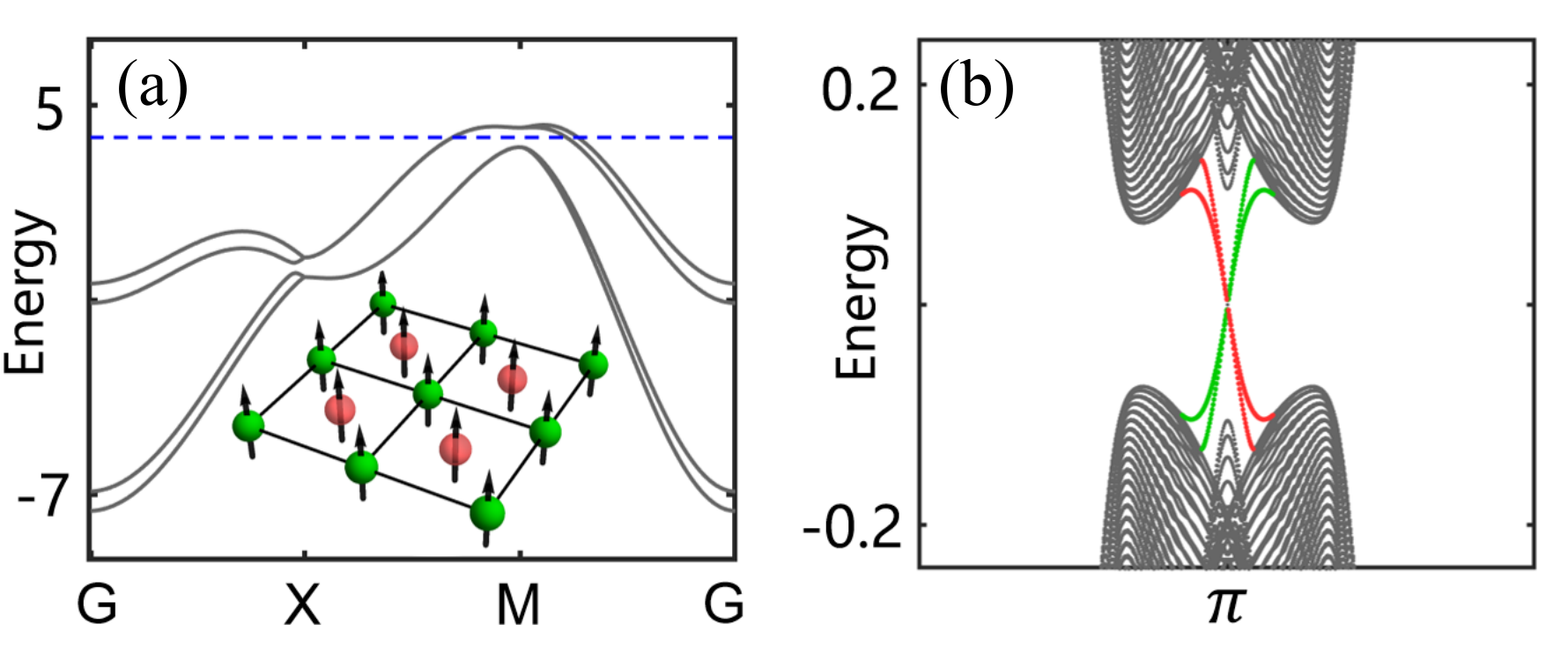}
	\caption{\label{fig4} Lattice structure, band structure and corresponding edge modes in the FM case.
  (a) The bands in the presence of the FM order with $\Delta_{\text{FM}} = 0.3$.
    The inset in (a) illustrates the real-space configuration of the FM order.
    The blue dashed line in (a) represent the chemical potential at $\mu = 4.0$.
    (b) shows the superconducting edge modes on the $(10)$ edge corresponding to the bands in (a), with the edge modes on the right (left) edge marked by the red (green) color.
    In (a) and (b), the other parameters are the same with these in Fig~\ref{fig2}.}
\end{figure}

\subsection{FM order induced first-order TSCs.}
We also consider FM order in the system in Eq.~\eqref{total_BdG}, and we assume the magnetic polarization along the $z$ direction. Correspondingly, the whole system can be depicted by the following Hamiltonian
\begin{equation}\label{main_FM}
\begin{aligned}
\mathcal{H}_{\text{BdG,FM}} = \mathcal{H}_{\text{BdG}} + \Delta_{\text{FM}} s_3 \sigma_0 \kappa_0,
\end{aligned}
\end{equation}
where $\Delta_{\text{FM}}$ is the strength of the FM order. We first study how the FM order affects the fourfold degenerate fermion at M in the normal state.
According to the real-space configuration in Fig.~\ref{fig4}(a), the symmetry of the system is lowered to the type-III magnetic space group $P4/n m^\prime m^\prime$ ($\#.129.417$)
\begin{equation}\label{group_sturcture_FM}
\begin{aligned}
\mathcal{G}_{\text{FM}}/T &= S_4^D \otimes Z_2 \otimes \{ \{ E | {\bf 0} \}, \{ M_y | {\bf 0} \} \mathcal{T} \},
\end{aligned}
\end{equation}
with $S_4^D$ the double group generated by $\{ S_{4z} | {\bf 0} \}$.
All the symmetry operations in $\mathcal{G}_{\text{FM}}/T$ maintain at the M point. A direct group theory analysis shows that, the 4D IR in the paramagnetic state at M splits into two 2D IRs similar to the AFM case.
However, differently in the FM case one corresponds to twofold band degeneracy between the $J_z = 1/2$ and $J_z = -3/2$ states and the other between the $J_z = -1/2$ and $J_z = 3/2$ states (more details in Supplementary Note 3).
Such degeneracies can be understood from the anticommutation relation between $\{ S_{4z} | {\bf 0} \}$ and $\{ I | {\bf \tau}_0 \}$ at M proved in Eq.~\eqref{main_129commute}. We confirm the above analysis numerically in Fig.~\ref{fig4}(a).

To study the topological property in systems depicted by $\mathcal{H}_{\text{BdG,FM}}$ corresponding to the normal bands in Fig.\ref{fig4}(a), we first calculate the Chern number.
Based on the formula in Eq.~\eqref{Chern_formula} and the above analysis, the Chern number can be calculated to be $| C_h | = 2$ whose sign depends on the sign of $\Delta_{{\text{FM}}}$.
To verify this, we simulate the superconducting edge modes numerically.
As shown in Fig.~\ref{fig4}(b), two chiral modes appear on each edge corresponding to the normal state in Fig.~\ref{fig4}(a), which is consistent with the above analysis.
In fact, the above chiral TSC state arises through a gap-close-reopen process at M as the FM order becomes stronger, and the phase transition occurs at $(4t + \mu)^2 + \Delta_{{\text{sc}}}^2 = \Delta_{{\text{FM}}}^2$.
Accordingly, in the weak-pairing condition the system is a TSC with $| C_h | = 2$, as long as the chemical potential is in the FM gap (details in Supplementary Note 4).
Moreover, the vortex in the chiral TSC state can also bind two Majorana modes, and the analysis is similar as that of the AFM case.
We present more detailed analysis and simulate the vortex bound states numerically in the Supplementary Note 8.

\begin{table}[t]
\caption{\label{Summary table}  Summary table for the $J_z$ of occupied band, the roles played by symmetries and the corresponding protected topology in different cases.}
\renewcommand\arraystretch{1.2}
\begin{tabular}{ p{0.7cm}<{\centering}p{1.64cm}<{\centering}p{1.5cm}<{\centering}p{2.5cm}<{\centering}p{1.7cm}<{\centering} }
\hline\hline
          &$J_z$ & Symmetry & Topology& Classification \\ \hline
       \multirow{2}{*}{AFM} & \multirow{2}{*}{$-1/2,-3/2$} & $\mathcal{M}_y\mathcal{T},\mathcal{P}$& Winding number & $\mathbbm{Z}$\\\cline{3-5}
       & & $S_{4z}$ &Fragile. Wan. Obs. & $\mathbbm{Z}_2$\\\hline
       FM&$+1/2,-3/2$&$\smallsetminus$ &Chern number &$\mathbbm{Z}$\\
\hline\hline
\end{tabular}
\end{table}

\section{Discussion}
We discuss the effects of the symmetry breaking perturbations \cite{footnote3}, which may arise from tilting the magnetization off the $z$ direction in Fig.~\ref{fig3}(a) and Fig.~\ref{fig4}(a), on the above TSC states.
Obviously, the vortex bound Majorana modes are sensitive to the $\{ S_{4z} | {\bf 0} \}$ breaking perturbations and will gap out immediately.
However, the Majorana edge and corner modes can persist against the perturbations. The chiral TSC state is robust as long as the bulk energy gap is not closed.
For the second-order TSC state, perturbations breaking $\{ M_y | {\bf 0} \} \mathcal{T}$ gap out the helical Majorana mode on the $(10)$ edge and break the $\mathbbm{Z}$ classification of the corner Majorana modes to a $\mathbbm{Z}_2$ one. Nevertheless, the corner Majorana modes can be more robust due to a $S_{4z}$ protected filling anomaly, or due to the boundary obstruction \cite{PhysRevX.10.041014,wong2023higher}.


In the proposal, the FM order can be replaced by an external magnetic field. More difficult is to construct the antiferromagnetic heterostructures, which require well-matched lattices between the magnetic layer and the layer offering the band degeneracy. A possible candidate is the heterostructure between the antiferromagnetism $A$Co$_2$As$_2$ ($A =$ Ca, Ba, Sr) and the iron-based superconductors, whose lattice constants are similar \cite{RevModPhys.87.855, PhysRevB.98.115128}. A more feasible scheme lies in the magnetic materials. For example, in Eu$_{1-x}$La$_x$FeAs$_2$ \cite{EuFeAs2} and Sr$_2$VO$_{3-\delta}$FeAs \cite{PhysRevB.105.214505}, magnetic layers exist next to the superconducting FeAs layer; and in Ba$_{1-x}$Na$_x$Fe$_2$As$_2$ \cite{BaNaFeAs} and Ba$_{1-x}$K$_x$Fe$_2$As$_2$ \cite{BaKFeAs}, a tetragonal AFM phase may coexist with the superconductivity. By methods of doping or gating, one may tune the chemical potential in the iron based superconductors near the fourfold band degeneracy and topological superconductivity can be possibly realized. We use the genuine bands of the iron-based superconductors to simulate the topological superconductivity in the Supplementary Note 9.

In the above analysis, we have mainly focused on the TSC states in the space group $P4/nmm$ and the possible material realization. However, as pointed out our method can be applied to a wide range of systems with band degeneracy near the Fermi energy. To further demonstrate the effectiveness of our method, we analyze another case where the lattice respects the space group $P4/ncc$. The group protects an eightfold band degeneracy at $(\pi,\pi,\pi)$ in the normal state. When conventional superconductivity is introduced, both the FM order and the C-type AFM order drives the system into the nodal TSC states, but the topological properties are different. More detailed analyses are presented in the Supplementary Note 10.
Another interesting point worth mentioning is that the symmetry of the system in the presence the magnetic order is determined by both the type of the magnetic order and the direction of the spin polarization; And it is possible that the higher-order TSC states can be realized by the simpler FM order, which deserves further study in the future.


In summary, we propose a general method which is based on the bulk energy bands and the conventional superconductivity to realize topological superconductivity.
We show that by manipulating systems with crystal symmetry-protected fermioic excitations with magnetism, TSCs including the high-order ones can be generally obtained when conventional superconductivity is introduced, and the property of the TSCs is thoroughly determined by the property of the magnetism.
Thus, our study provides a new method to realize the various types of topological superconductivity and can help to find new platforms to realize the Majorana modes.

Near finishing the paper, we become aware of a work \cite{vortex_second} in which the vortex bound states in high-order TSCs are studied, and the conclusion in the work is consistent with our results in the second-order TSC state in the AFM case.

\section{Methods}

{\bf {Symmetries in time-reversal invariant superconductors.}}
Generally, a superconductor can be described by the following BdG Hamiltonian
\begin{equation}\label{eq_HTR1}
  \mathcal{H}_{\mathrm{BdG}}(\kk)=\begin{pmatrix}
    \mathcal{H}_0(\kk)-\mu&\Delta(\kk)\\
    \Delta^\dagger(\kk) &-\mathcal{H}_0^\ast(-\kk)+\mu\\
  \end{pmatrix},
\end{equation}
in the basis $\psi^\dagger({\bf k}) = ( c^\dagger_{{\bf k}, \uparrow}, c^\dagger_{{\bf k}, \downarrow}, c_{-{\bf k}, \uparrow}, c_{-{\bf k}, \downarrow} )$. Notice that we neglect other indices except for the spin index here. For a time-reversal symmetric superconductor, it respects the following three symmetries: the time-reversal symmetry $\mathcal{T}$, the particle-hole symmetry $\mathcal{P}$ and the combined chiral symmetry $\mathcal{C} = \mathcal{P} \mathcal{T}$. These symmetries act on the Hamiltonian as follows
\begin{equation}\label{eq_PTC}
\begin{split}
  \mathcal{T} \mathcal{H}_{\mathrm{BdG}}(\kk) \mathcal{T}^{-1}&=\mathcal{H}_{\mathrm{BdG}}(-\kk), \\
  \mathcal{P} \mathcal{H}_{\mathrm{BdG}}(\kk) \mathcal{P}^{-1}&=-\mathcal{H}_{\mathrm{BdG}}(-\kk),\\
  \mathcal{C} \mathcal{H}_{\mathrm{BdG}}(\kk) \mathcal{C}^{-1}&=-\mathcal{H}_{\mathrm{BdG}}(\kk).
\end{split}
\end{equation}
Moreover, in the basis for $\mathcal{H}_{\mathrm{BdG}}(\kk)$ in Eq. \eqref{eq_HTR1}, the above symmetries take the form $\mathcal{T}=is_2 \kappa_0 K$, $\mathcal{P} = s_0 \kappa_1 K$ and $\mathcal{C} = is_2 \kappa_1$.
Besides the above local symmetries, the system also respects the crystalline symmetries. The crystalline symmetry $\Tilde{g}$ transforms the BdG Hamiltonian as $\Tilde{g} \mathcal{H}_{\mathrm{BdG}}(\kk) \Tilde{g}^{-1} = \mathcal{H}_{\mathrm{BdG}}(\Tilde{g}^{-1} \kk)$, and has the form
\begin{equation}\label{eq_crystallineSSQ}
  \Tilde{g}=\begin{pmatrix}
    g & 0\\
    0 & \eta g^\ast\\
  \end{pmatrix}.
\end{equation}
In the above equation, $\eta$ is determined by the pairing symmetry, i.e. $g \Delta(\kk) g^T = \eta \Delta(\kk)$. In the present study, we focus on the conventional superconductivity, which belongs to the trivial irreducible representation of the crystalline symmetry group. Namely, $\eta$ always equals 1 for $\Tilde{g}$ in Eq. \eqref{eq_crystallineSSQ} in our consideration.

Then, we consider the commutation relation between the unitary chiral symmetry $\mathcal{C}$ and the crystalline symmetries. It can be directly shown
\begin{equation}\label{eq_commuteSSQ}
  \mathcal{C} \Tilde{g} \mathcal{C}^{-1} = \begin{pmatrix}
    s_2 g^\ast s_2 & 0 \\
    0 & s_2 g s_2 \\
  \end{pmatrix},
\end{equation}
where we have taken use of the fact $\eta = 1$ in $\Tilde{g}$. Recall that in a time-reversal symmetric system the time-reversal symmetry commutes with all the crystalline symmetries, and in the normal state it demands $T g T^{-1} = g = (is_2 K) g (is_2 K)^{-1} = s_2 g^\ast s_2$ where $T$ stands for the time-reversal symmetry in the normal state. Therefore, we have $\mathcal{C} \Tilde{g} \mathcal{C}^{-1} = \Tilde{g}$ in Eq. \eqref{eq_commuteSSQ}, namely $[\mathcal{C}, \Tilde{g}] = 0$. The above commutation relation leads to that for any eigenstate $| \phi(\kk) \rangle$ of $\mathcal{H}_{\mathrm{BdG}}(\kk)$ carrying energy $E(\kk)$, its chiral partner $\mathcal{C} | \phi(\kk) \rangle$ possesses energy $-E(\kk)$ but the same symmetry eigenvalue with $| \phi(\kk) \rangle$ for any crystalline symmetry. This means that in the level of the symmetry indicator, the system must equal to the topological trivial superconductor.
More detailed analyses are presented in Supplementary Note 1.

{\bf {The calculation of winding number.}}
To analytically calculate the winding number at $k_y=\pi$ in the AFM case, we rewrite $\mathcal{H}_{\text{BdG,AFM}}$ in Eq.~\eqref{main_AFM} in the basis diagonalizing the pseudo-chiral symmetry $\widetilde{\mathcal{C}}$.
After the basis transformation, $\mathcal{H}_{\text{BdG,AFM}}$ takes an off-diagonal form in the Nambu space. In the specific AFM case, the off-diagonal block matrix $Q(k_x)$ in the upper right corner is
\begin{equation}
\begin{split}
Q(k_x)=\begin{pmatrix} q_+(k_x)&0\\0&q_-(k_x) \end{pmatrix},
\end{split}
\end{equation}
with $q_{\pm}(k_x) = -[2t(\cos k_x-1)-\mu]s_0\pm2\lambda \sin k_x s_2\pm\Delta_{\text{AFM}} s_3 \pm i \Delta_{\text{sc}} s_2$.
Accordingly, the winding number along $k_y = \pi$ can be calculated as
\begin{equation}
\begin{aligned}
\nu(Q) &= \frac{i}{2\pi}\int_{0}^{2\pi} d k_x \partial_{k_x} \log [\det Q(k_x)]\\
&= \frac{i}{2\pi}\int_{0}^{2\pi} d k_x \partial_{k_x} \log [(\det q_+(k_x)\det q_-(k_x))]\\
&= \frac{i}{2\pi}\int_{0}^{2\pi} d k_x \partial_{k_x}[\log(\det q_+)+\log(\det q_-)]\\
&= \nu(q_+) + \nu(q_-).
\end{aligned}
\end{equation}
Here, $\nu(q_\pm)$ characterizes the winding number of $\det q_\pm(k_x)$ around the origin point in the complex plane.
See more details in Supplementary Note 4.

{\bf {A short review of MEBR.}}
When we place the bases $\{\phi_i^\alpha\}$ of the irreducible co-representations $u_i$ of these on-site magnetic point groups $\mathcal{G}_{\mathbf{x}}$ at their corresponding Wyckoff positions $\mathbf{x}$, the induced co-representation $(u_i)_{\mathbf{x}}\uparrow \mathcal{G}$ of the space group $\mathcal{G}$ from the irreducible co-representations of the subgroup $\mathcal{G}_{\mathbf{x}}$ is referred to as magnetic elementary band representations (MEBR).
In the AFM case, the four negative energy bands host the co-representations
\begin{itemize}
  \item At G point: $\bar{\Gamma}_7\oplus\bar{\Gamma}_7$
  \item At M point: $\bar{M}_7\oplus \bar{M}_7$
  \item At X point: $\bar{X}_3\bar{X}_5\oplus \bar{X}_2\bar{X}_4$
\end{itemize}
Therefore, our target band can only be expressed as a combination of MEBRs with the negative integer
\begin{equation}
    (\bar{E})_{2c}\uparrow \mathcal{G}_{\text{AFM}}\ \oplus\ ( ^1 \bar{E}_1)_{2a}\uparrow \mathcal{G}_{\text{AFM}} \ \circleddash \ ( ^1 \bar{E}_2)_{2a}\uparrow \mathcal{G}_{\text{AFM}},
\end{equation}
which implies the fragile topology. See more details in Supplementary Note 7.

{\bf {Model Hamiltonian used for SG 130.}}
To illustrate the effectiveness and generality of our method, we introduce a more complex example for space group $P4/ncc$ ($\#.130$). We start with the paramagnetic normal state, where the system actually respects the type-II magnetic space group $P4/ncc1^\prime$. The group can be generated by the following symmetry operations
\begin{equation}
    \{C_{4z}|000\},\ \{C_{2x}|\frac{1}{2}\frac{1}{2}0\},\ \{I|\frac{1}{2}\frac{1}{2}\frac{1}{2}\},\ \mathcal{T}.
\end{equation}
The magnetic space group $P4/ncc1^\prime$ has one and only one eightfold irreducible representation at the $A$ point, i.e. the $(\pi, \pi, \pi)$ point, in the spinful condition. Namely, all the bands are eightfold degenerate and respect the same low-energy effective model in the spinful case. In the lattice model condition, the eightfold band degeneracy can be captured by the following tight-binding model\cite{PhysRevLett.116.186402}
\begin{equation}
\begin{split}
    \mathcal{H}_0(\kk)&=t_0(\cos k_x+\cos k_y+\cos k_z)+t_{xy}\tau_x\cos\frac{k_x}{2}\cos\frac{k_y}{2}\\
    &+t_z\mu_x\cos\frac{k_z}{2}+\lambda_1\tau_z\mu_y\cos\frac{k_z}{2}\\
    &+\lambda_3\tau_x\mu_z\left(\sigma_x\sin\frac{k_x}{2}\cos\frac{k_y}{2}+\sigma_y\cos\frac{k_x}{2}\sin\frac{k_y}{2}\right)\\
    &+\lambda_2\tau_z(\sigma_x\sin k_y-\sigma_y \sin k_x).
\end{split}
\end{equation}
Based on this model we study the possible TSC states in the system, when conventional superconductivity and different magnetic orders are introduced.
More details are presented in Supplementary Note 10.

\subsection{Data availability} All data needed to evaluate the conclusions in the study are present in
the paper and/or the Supplementary Information. The data that support the findings of this study are available from the corresponding authors upon request.

\subsection{Code availability} The computer code used for numerical calculation and theoretical understanding is available upon request from the corresponding authors.

\subsection{References}
\bibliography{ref}

\begin{thebibliography}{78}
\bibitem [{foo({\natexlab{a}})}]{footnote2}%
  \BibitemOpen
  \href@noop {} {\bibfield  {journal} {\bibinfo  {journal} {In fact, in the
  nonsymmorphic group $P4/nmm$ besides the $S_{4z}$ symmetry, there is also the
  $C_{4z}$ symmetry which is defined at the center of the square formed by the
  four nearest neighbouring lattice sites in Fig.~\ref{fig2}(a), and we have
  specified this point in the SM. As group $P4/nmm$ merely has one 4D IR at M,
  all the analyses related to $S_{4z}$ also work for $C_{4z}$}\ }
  ({\natexlab{a}})}\BibitemShut {NoStop}%
\bibitem [{foo({\natexlab{b}})}]{footnote3}%
  \BibitemOpen
  \href@noop {} {\enquote {\bibinfo {title} {{For the TSC states, more
  essential are the symmetries in the magnetic states rather than the specific
  magnetic orders considered in Fig.~\ref{fig3}(a) and Fig.~\ref{fig4}(a).
  Therefore, here we refer to the perturbations breaking the magnetic group
  symmetries}},}\ } ({\natexlab{b}})\BibitemShut {NoStop}%
\end{thebibliography}

\subsection{Acknowledgments}
This work is supported by the Ministry of Science and Technology (Grant No. 2022YFA1403901 and NO. 2022YFA1403902), National Natural Science Foundation of China (Grant NO. NSFC-12304163, NSFC-12325404, and NSFC-11920101005), National Key R$\&$D Program of China (Grant NO. 2022YFA1403800 and NO. 2023YFA1406700), Innovation program for Quantum Science and Technology (Grant No. 2021ZD0302500), Chinese Academy of Sciences (Grant NO. XDB33020000 and NO. JZHKYPT-2021-8), the New Cornerstone Investigator Program, and the Beijing Institute of Technology Research Fund Program for Young Scholars. Z. Wu and Y. Wang are supported by NSF under award number DMR-2045781.



\subsection{Author contributions}
Z.Y.Z., Z.F.W., Y.W and S.S.Q. did the theoretical derivation and numerical calculation; C.F., F.-C.Z., J.P.H, Y.W., S.S.Q. provided the theoretical understanding. All authors discussed and contributed to the manuscript. S.S.Q. and Y.W conceived the work.

\subsection{Competing interests}
The authors declare no competing interests.

\end{document}